\documentclass{article}

\usepackage{arxiv}
\usepackage{graphicx}
\usepackage[utf8]{inputenc} 
\usepackage[T1]{fontenc}    
\usepackage{hyperref}       
\usepackage{url}            
\usepackage{booktabs}       
\usepackage{amsfonts}       
\usepackage{nicefrac}       
\usepackage{microtype}      
\usepackage{lipsum}

\title{STOCHASTIC SPREAD PAIRS TRADING IN THE INDIAN COMMODITY MARKET}
\author{
    Dhruv Mahajan\thanks{} \\
  Department of Mining Engineering\\
  IIT Kharagpur\\
  Kharagpur,WB,721302 \\
  \texttt{dhruvmahajan4291@iitkgp.ac.in} \\
   \And
 Abhijeet Chandra \\
  Vinod Gupta School Of Management\\
  IIT Kharagpur\\
  Kharagpur,WB,721302 \\
  \texttt{abhijeet@vgsom.iitkgp.ac.in } \\
}

\begin{document}
\maketitle
\begin{abstract}
In this study, we applied a stochastic spread pairs trading strategy on the Indian commodity market. The complete set of commodities were taken whose spot price was available for the period of January 1st 2010 to December 31st 2018 including energy, metals and the agricultural commodity sector. Spot data was taken from the MCX pooled spot prices for 17 commodities. The data was split into training period (January 1st 2010 to 14th March 2017) and testing period(15th Match 2017 to 31st December 2018). The splitting was done using a 80:20 split.Johanssen Cointegration tests were done on training data for pairs of commodities to check for long-run relationship and the cointegrated commodities were selected for formation of the trading process. We found a total of 12 cointegrated pairs out of 136 possible pairs. Cointegration was assumed for the testing period. A single-factor stochastic trading approach was applied on the logarithmic spread of the cointegrated pairs for both the training and testing period.The parameters of stochastic spread model were estimated using differential evolution algorithm. Also parameters for the trading rule were optimized by backtesting on the training period and assumed for the testing period. The results show a sharpe ratio of above 1.4 for all the commodity cointegrated pairs in the backtesing period.  
\end{abstract}


\section{Introduction}
Pairs trading is a statistical arbitrage trading strategy that is used to profit from diversions from the equilibrium state created in the spread process of two cointegrated assets. It involves buying(going long) on the over-valued asset and selling(going short) on the underlying assets. “Longing/Shorting the spread” is also used to describe the trade.

Pairs trading is credited to Gerry Bamberger,David Shaw and Nunzio Tartaglia who studied arbitrage opportunities in the equity markets in the mid-1980s (Latte, T. (2011)). In 1987, the strategy made Morgan Stanley and Co. A reported 50 million USD in profits. Over the years, pairs trading has been used by various hedge funds and investment mangers and continues to be one of the most widely used statistical arbitrage trading strategy. 

The main essence behind pairs trading is that a spread created as a linear combination of prices or log prices of two cointegrated securities is mean reverting in nature i.e the spread in the long run stays around an equilibrium level. When the spread diverges from this equilibrium level, either in the upward or downward direction, the trader takes a short or a long position respectively assuming the spread will revert to the mean after a period of time. This same idea can be extended to commodities. There is a tendency for prices of some commodities to move in the same direction, this is a result of common factor exposures of commodity prices to demand and supply shocks which affects their prices in a similar fashion(Labys et al. 1999). Another reason is that some metals are produced together from mining operations thus their prices are affected in a similar fashion.

Two series are said to be cointegrated if a linear combination of these series is stationary.The cointegration of the log price series of two commodities is tested using the Johanssen Cointegration test. The johansen test includes a Vector Error Correction Model(VECM) by first order differencing of an VAR(p) series. The rank of the first coffecient matrix help in determining the hypothesis of cointegration.Eigenvalue decomposition of the first coffecient matrix is carried out which results in a series of eigenvectors. The components of the eigenvector with largest eigenvalue form the hedge ratio.

This paper contributes to the existing literature by testing a stochastic spread pairs trading strategy on the Indian Commodity market. The stochastic model assumes that the spread is driven by a latent state variable which follows the mean reverting Ornstein-Uhlenbeck process(popularly known as the vasicek model in finance). The spread is equal to the latent variable plus a multiple of gaussian noise with mean zero and variance one (Elliott et al. (2005)) . This two equation formulation is modelled using a space state model.There has been a very limited amount of reseach done regarding profitability of stochastic pairs trading models both in the equity and commodity markets. We also propose a more stable optimization routine for the parameter estimation of the stochastic model. Instead of Expectation Maximization (EM) algorithm used in Elliott et al. (2005), we propose a maximum likelihood estimated of parameters of the two equations optimized using the Differental Evolution algorithm, a kind of evolutionary algorithm.

The results show that in the backtesting space, most of the pairs generated a positive cagr with a sharpe ratio of greater than 1.4. This implies that trading in commodities using pairs trading can be a profitable investment in Emerging markets like India where not much quantitative trading happens in the commodity sector.In the testing space, some of the pairs had negative sharpe ratio which can be accounted to regime shifts in the spread. Detailed results are given in the ‘Results’ section.

\section{Literature Review}

\section{Data and Methodology}
\label{sec:headings}

\subsection{Data}
The data we used consists of daily logarithmic prices of the following commodities: 
\linebreak
  \linebreak
    Aluminium
    \linebreak
    Almond
    \linebreak
    Lead
    \linebreak
    Cardomom
    \linebreak
    Crude
    \linebreak
    Gold
    \linebreak
    Copper
    \linebreak
    Natural-Gas
    \linebreak
    Zinc
    \linebreak
    Wheat
    \linebreak
    Tin
    \linebreak
    Silver
    \linebreak
    Soyoil
    \linebreak
    Jute
    \linebreak
    Nickel
    \linebreak
    Guargam
    \linebreak
    Cpo

We considered data for the all the commodities for whom daily spot prices were available from MCX website for the period of  January 1st 2010 to 31st December 2018. To have a consistent series across all commodities, we filled the missing values for dates using linear interpolation of data of nearby dates for that commodity. 

The data was divided into two sets: Back testing(Training) set and the Testing set. There is no formal rule of splitting between the two sets.  The training period needs to be long enough to test for co integration and optimize various parameters of the stochastic model, but not so long that there is not enough information left for the analyze for the testing period. The testing period needs to be long enough to test our strategy on out of sample data but it cannot be too long because it is possible that the co integration relationship between two tested commodities may change. The 10 year data consisted of a total of 2347 observations with 1878 observations in the training set and 469 observations in the testing set.

\subsection{Cointegration}

The main idea behind pairs trading is that two time series are related to each other in the long run, meaning their long term price spreads are in equilibrium and will mean revert back to this equilibrium state a short while after the state is disturbed. This long term relationship can be tested using co integration tests.

The idea is to find two non-stationary time series and using a linear transformation on these two series, produce a stationary series. The fact that the input series need to be non-stationary is due to the fact that linear combination of two stationary series will always be stationary and hence the cointegration test will go void.

$ Let \ x_t \ and  \ y_t \ be \ two \ non-stationary\ time \ series\ ,if \  the \ linear \ combination\ ax_t + by_t\ is\ stationary\ then\ we \ say \ that \ x_t \ and \ y_t \ are \ co- integrated.$

 Where $\ a, b  \in R $ , in our case the ratio $b/a$ would give us the hedge ratio.

To test for co-integration,we used the Johansen's co-integration test, details for which are given in the next subsection.
 
\subsubsection{Johansen Test}
To understand the details of how the test works,an introduction to Vector Auto-regressive Models(VAR) is required. A general form of VAR(p) model without the drift is given by:

\begin{equation}
{\bf y_t} = {\bf \mu} + A_1 {\bf y_{t-1}} + \ldots + A_p {\bf y_{t-p}} + {\bf w_t}  
\end{equation}

Where  $ \mu $ is the vector mean of the vector series $y_t$, $A_i$ are the coefficient matrices and $w_t$ is multivariate Gaussian noise.

A Vector Error Correction Model (VECM) can be defined by taking the first order different of the above VAR model.
\begin{equation}
\Delta {\bf y_t} = {\bf \mu} + A {\bf y_{t-1}} + \Gamma_1 \Delta {\bf y_{t-1}} +  \ldots + \Gamma_p \Delta {\bf y_{t-p}} + {\bf w_t}
\end{equation}

Where $\Delta y_t = y_t - y_{t-1} $, $A$ is the coefficient matrix for $y_{t-1}$ and $\Gamma_i$ are the matrices for each difference lag.

The Johansen test checks for the rank $r$ of matrix $A$. In case of pairs, $A$ will be a $2X2$  matrix and the hypothesis for the test will be:

$H0 : r = 0 $

$H1 : r \neq 0$

The null hypothesis implies no co-integration, and the rejection of the null hypothesis implies co-integration.

To test for the hypothesis an eigenvalue decomposition of $A$ is carried out which results in a series of eigenvectors. The components corresponding to the largest eigenvector in case of rejection of null hypothesis gives us the linear combination required for a stationary series, in other words the ratio $b/a$ or the hedge ratio.

\subsubsection{Co-integration Results}

Out of a total of 136 possible pairs from 17 commodities, we found a total of 12 co-integrated pairs: 

\begin{center}
 \begin{tabular}{||c c c||} 
 \hline
 Commodity1 & Commodity2 & Hedge Ratio \\ [0.5ex] 
 \hline\hline
 Aluminium & Lead & $-0.482$ \\ 
 \hline
 Aluminium & Tin & $-0.339$ \\
 \hline
 Aluminium &  Soyoil & $-0.236$  \\
 \hline
 Aluminium & CPO & $-0.242$  \\
 \hline
 Lead & Gold & $-0.560$ \\ 
 \hline
 Lead & Natural Gas & $-0.619$ \\ 
 \hline
 Lead & Wheat & $-0.695$ \\ 
 \hline
 Lead & Tin & $-0.788$ \\
 \hline
 Lead & Soyoil & $-0.634$ \\
 \hline
 Gold & Soyoil & $-1.157$ \\
 \hline
 Natural Gas & Tin & $-1.325$ \\
 \hline
 Silver & Guargum & $-0.319$ \\
 \hline
\end{tabular}
\end{center}

\subsubsection{Reasons for Co-integration}

\subsection{Stochastic Spread Model}

The spread between two co-integrated variables is assumed to be driven by a latent state variable $\zeta$ which follows a Ornstein-Uhlenbeck stochastic process (Elliott et al. (2005)) 

\begin{equation}
    d\zeta_t = \kappa(\mu - \zeta_t)dt + \sigma dB_t
\end{equation}

And the spread between stocks $i$ and $j$ : $s_{t,i,j}$ is driven by the state variable $\zeta$ and Gaussian noise $w_t$ with a mean $0$ and variance $1$.

\begin{equation}
    s_{t,i,j} = \zeta_t + Vw_t
\end{equation}

Where $\kappa$ is speed of mean reversion or we can say that $1/\kappa$ gives us the half-life of mean reversion, that is the number of days we can expect a single pair trade to last.

$\theta$ is the long-term mean of the spread, around which we can expect the series to be in an equilibrium state,whenever the spread deviates from this mean we will either 'long the spread' or 'short the spread'

$B_t$ is a standard Brownian motion. $V$ gives us the assumed constant variance of the noise term.

Using the fact that solution to equation 3 is markovian, it can be converted to a discrete time transition equation:

\begin{equation}
    \zeta_{t} = X + Y\zeta_{t-1} + Z\epsilon_{t}
\end{equation}

And the measurement equation is nothing but equation 4:

\begin{equation}
    s_{t,i,j} = \zeta_t + Vw_t
\end{equation}

Where $\epsilon_t$ is an error term such that $\epsilon_t \sim \mathcal{N}(0,1) $

For a time interval of 1-day, that is daily observations we can make the following conclusions:

\begin{equation}
    X = \mu / \kappa
\end{equation}

\begin{equation}
   Y = 1-\kappa
\end{equation}

\begin{equation}
   Z = \sigma
\end{equation}

Where $X$ $Y$ and $Z$ follow the following constraints:

\begin{equation}
    X \geq 0
\end{equation}

\begin{equation}
    0 < Y < 1  
\end{equation}

\begin{equation}
   Z \geq 0 
\end{equation}

Equations 5 and 6 can be seen as a pair of linear space state equations with former being the transition equation and latter being the measurement equation. Hence they can be modelled using Kalman Filter. Kalman Filtering and Smoothing consists of broadly two steps:

First step is filtering, which is estimation of  $\zeta_t$ given $D_t$ where $D_t$ is the information available at time $t$ that is $[s_1,s_2....s_t]$

Second step is smoothing, which is to estimate $\zeta_t$ given $D_T$ where $T>t$

The detailed theory behind Kalman Filters is beyond the scope of this paper, but it is important to know that there is a likelihood value corresponding to every such model.

\subsection{Parameter Estimation}

The maximum likelihood estimation of $X , Y, Z and V$ is done using a Differential Evolution algorithm subject to the constraints 10,11 and 12. Differential Evolution is an instance of Evolutionary Algorithm from the field of Evolutionary Computation. 

The Differential Evolution algorithm involves iterations of :Recombination, Evaluation, and Selection applied to the population of possible solutions. The recombination approach involves the creation of new candidate solution components based on the weighted difference between two randomly selected population members added to a third population member.

Many of the advantages of Differential Evolution is its ability to optimize functions that are non-differentiable, non-continuous, non-linear, noisy, flat, multi-dimensional or have many local minima, constraints or stochasticity. But the most notable advantage is that it does not require an initial set of variable to set as the solution, it only requires a range of possible solutions, over which the algorithm search for the optimum value.

The parameters in this case are optimized in a way that the log-likelihood value given by the Kalman filer is maximized.
Assuming $X^* , Y^*, Z^*$ and $V^*$ be the optimal parameters, then we can find out the spread parameters as:

\begin{equation}
    \kappa = 1 - Y^*
\end{equation}

\begin{equation}
    \mu  = X^*/\kappa
\end{equation}

\begin{equation}
   \sigma = Z^*
\end{equation}

\subsection{Trading Strategy}

Using the main idea behind pairs trading, that is entering into a trade whenever it deviates from it's equilibrium state, we can design a trading strategy centred around $\mu$ and design a set of rules for the spread $s_{t,i,j}$  for entering and leaving from a trade.Elliott et al. (2005) provides such a trading rule. 

Short the spread when:
\begin{equation}
    s_{t,i,j} \geq \mu + c_{i,j}(\sigma/\sqrt{2\kappa})
\end{equation}

Long the spread when 

\begin{equation}
    s_{t,i,j} \leq \mu - c_{i,j}(\sigma/\sqrt{2\kappa})
\end{equation}

Exit Short trade when

\begin{equation}
    s_{t,i,j} \leq \mu + 0.0001s_{t,i,j}
\end{equation}

Exit Long trade when
\begin{equation}
    s_{t,i,j} \geq \mu - 0.0001s_{t,i,j}
\end{equation}

The parameter $c_{i,j}$ is optimized using a vector search over the Backtesting period, the value of $c_{i,j}$ which optimizes the Sharpe Ratio of the trading strategy is chosen and is assumed constant over the Testing period.

\section{Empirical Results}

The results are divided into two sections, the backtesting results and the testing results, which gives us the in-the-sample and out-of-the-sample results respectively.

\subsection{Backtesting Results}

We calculated various metrics for our trading strategy including Sharpe Ratio, Cummalative Return, Average Return ($\mu_{return}$), Standard Deviation of returns ($\sigma_{return}$) ,Cumulative Annual Growth Rate (Cagr) , Maximum Drawdown, Skewness , Kurtosis and number of long and short trades.

Where,

\begin{equation}
   Sharpe  \ Ratio = (\mu_{return} - r_f)/\sigma_{return}
\end{equation}

$r_f$ is the risk free rate and is taken to be the yield of 10YR Indian Treasury Bond which is equal to $0.074$

\begin{center}
 \begin{tabular}{||c c c c c c c c c c||} 
 \hline
 $Commodity1$ & $Commodity2$ & $c_{i,j}$ & $SR$ & $CAGR$  & $Max \ Drawdown$ & $Skew$ & $Kurt$ & $N_{long}$ & $N_{short}$\\ [0.5ex] 
 \hline\hline
 Aluminium & Lead & $80.6$ & $1.46$ & $12.61\%$ & $-14\%$ & $0.60$ & $6.23$ & $39$ & $71$\\ 
 \hline
 Aluminium & Tin & $82.1$ & $1.50$ & $13.45\%$ & $-15\%$ & $0.37$ & $5.97$ & $34$ & $87$\\
 \hline
 Aluminium &  Soyoil & $0.28$ & $1.81$ & $15.90\%$ & $-19\%$ & $0.26$ & $2.16$ & $64$ & $64$\\
 \hline
 Aluminium & CPO & $133.03$ & $1.48$ & $12.03\%$ & $-14\%$ & $0.84$ & $10.48$ & $23$ & $17$\\ 
 \hline
 Lead & Gold &  $19.79$ & $1.54$ & $18.70\%$ & $-29\%$ & $2.41$ & $28.94$ & $36$ & $47$\\ 
 \hline
 Lead & Natural Gas & $26.69$ & $1.89$ & $21.76\%$ & $-30\%$ & $1.08$ & $11.22$ & $32$ & $32$\\ 
 \hline
 Lead & Wheat & $55.39$ & $1.41$ & $17.30\%$ & $-31\%$ & $1.28$ & $30.68$ & $36$ & $39$\\
 \hline
 Lead & Tin & $13.82$ & $2.03$ & $22.64\%$ & $-25\%$ & $2.15$ & $23.48$ & $54$ & $76$\\
 \hline
 Lead & Soyoil  & $24.08$ & $2.14$ & $21.50\%$ & $-27\%$ & $1.64$ & $20.17$ & $56$ & $47$ \\
 \hline
 Gold & Soyoil & $88.00$ & $1.95$ & $14.58\%$ & $-21\%$ & $0.38$ & $12.16$ & $16$ & $27$\\
 \hline
 Natural Gas & Tin & $26.41$ & $1.98$ & $25.53\%$ & $-40\%$ & $0.60$ & $6.23$ & $39$ & $71$\\
 \hline
 Silver & Guargum & $139.78$ & $1.02$ & $10.73$ & $-19\%$ & $0.73$ & $8.25$ & $7$ & $3$\\
 \hline
\end{tabular}
\end{center}

\subsection{Testing Results}

The following table provides the out of sample results

\begin{center}
 \begin{tabular}{||c c c c c c c c c c||} 
 \hline
 $Commodity1$ & $Commodity2$ & $c_{i,j}$ & $SR$ & $CAGR$  & $Max \ Drawdown$ & $Skew$ & $Kurt$ & $N_{long}$ & $N_{short}$\\ [0.5ex] 
 \hline\hline
 Aluminium & Lead & $80.6$ & $1.51$ & $46.10\%$ & $-38\%$ & $-3.50$ & $54.32$ & $0$ & $16$\\ 
 \hline
 Aluminium & Tin & $82.1$ & $1.00$ & $150.92\%$ & $-282\%$ & $-3.47$ & $53.30$ & $4$ & $10$\\
 \hline
 Aluminium &  Soyoil  & $0.28$ & $-$ & $0\%$ & $0\%$ & $0$ & $0$ & $0$ & $0$\\ 
 \hline
 Aluminium & CPO &  $133.03$ & $1.03$ & $114.13\%$ & $-193\%$ & $-1.84$ & $32.05$ & $0$ & $8$\\ 
 \hline
 Lead & Gold & $19.79$ & $-1.06$ & $-42.96\%$ & $-91\%$ & $-2.44$ & $102.71$ & $2$ & $1$\\
 \hline
 Lead & Natural Gas & $26.69$ & $1.58$ & $122.62\%$ & $-138\%$ & $0.67$ & $4.47$ & $3$ & $2$\\ 
 \hline
 Lead & Wheat & $55.39$ & $3.89$ & $67.35\%$ & $-27\%$ & $0.15$ & $9.11$ & $5$ & $6$\\
 \hline
 Lead & Tin & $13.82$ & $2.87$ & $78.99\%$ & $-51.7\%$ & $0.32$ & $3.18$ & $6$ & $9$\\
 \hline
 Lead & Soyoil  & $24.08$ & $-0.27$ & $-4.62\%$ & $-82\%$ & $0.17$ & $15.90$ & $2$ & $5$ \\
 \hline
 Gold & Soyoil & $88.00$ & $1.38$ & $29.60\%$ & $-15.4\%$ & $0.11$ & $1.92$ & $9$ & $4$\\
 \hline
 Natural Gas & Tin & $26.41$ & $0.20$ & $21.78\%$ & $-199\%$ & $2.45$ & $38.78$ & $5$ & $3$\\
 \hline
 Silver & Guargum & $139.78$ & $-$ & $0\%$ & $0\%$ & $0$ & $0$ & $0$ & $0$\\
 \hline
\end{tabular}
\end{center}

As we can see from the table most of the returns exhibit very high drawdown with a very high CAGR which point out to high volatility in testing returns, one possible reason for that might be regime shifts in spreads. Even with that, our strategy was able to gain a positive sharpe ratio in most of the cases with very high annual returns.

Below figures give the cummalative returns over the testing period for 3 commodity pairs.

\graphicspath{ {./images/} }

\includegraphics[scale = 0.35]{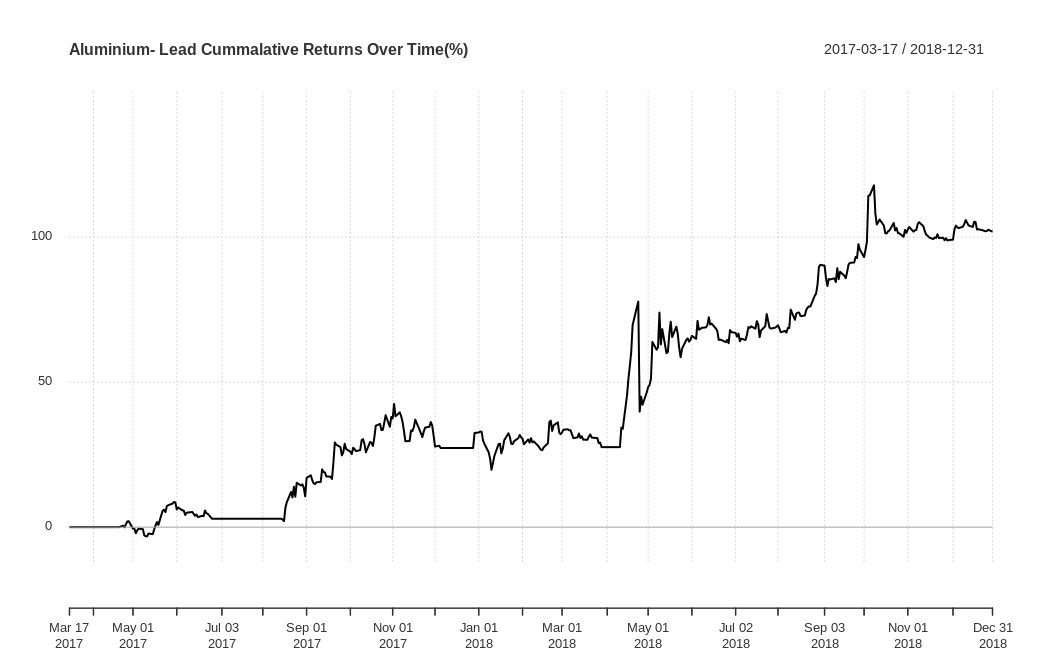}
\includegraphics[scale = 0.35]{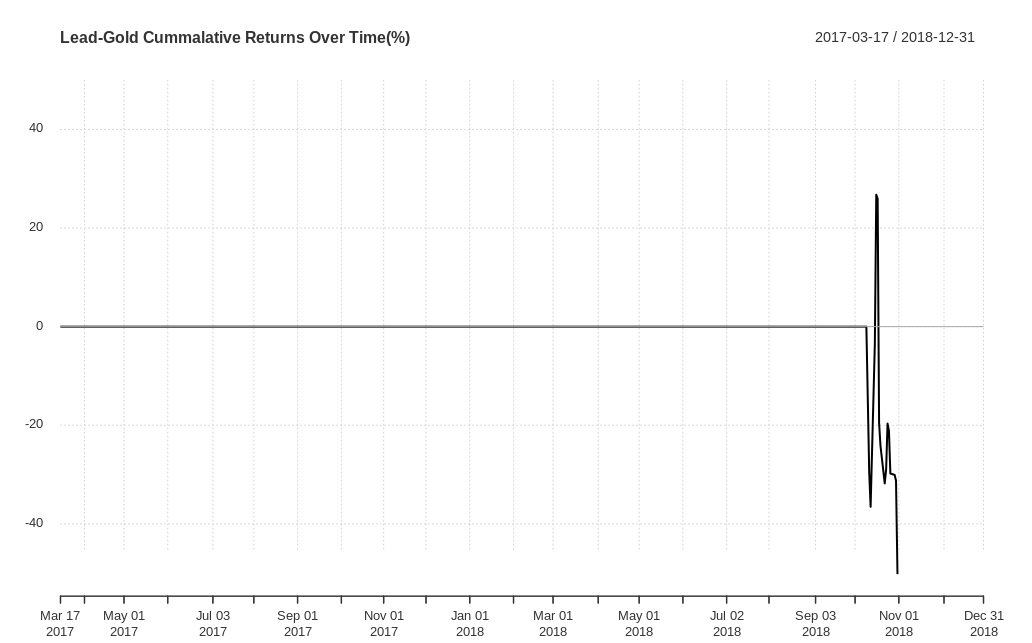}
\includegraphics[scale = 0.35]{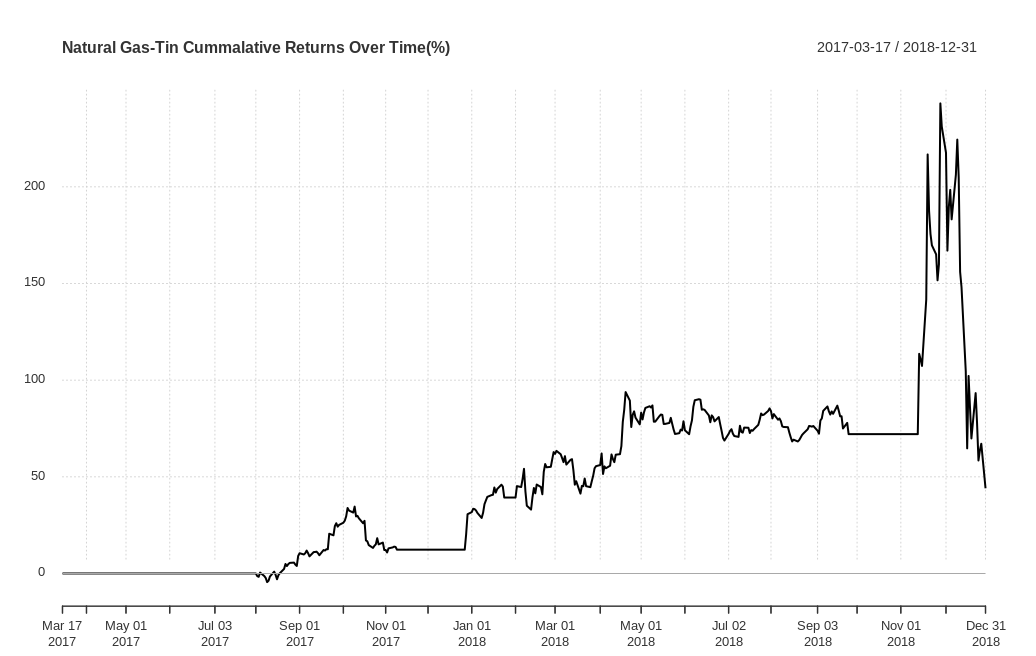}

\section{Conclusion}

We have applied a general stochastic approach to model relative mispricing between two commodities for a pairs trading setup. A novel approach to estimating the parameters for the stochastic model using a differential evolution algorithm is discussed. We applied co-integration tests to 17 different commodities, and out of 136 possible pairs, only 12 were found to be co-integrated. After optimizing for the various parameters we found that in the backtesting period we can get a Sharpe ratio as high as 2.14 with a CAGR of 21.5\% which is quite higher than the market returns during that period. Also during the testing period we found that the returns were quite volatile but nonetheless the strategy managed to produce overwhelming returns as high as 151\% for a pair and positive returns for most of the pairs.This paper does not address the source of such high volatility in the testing period.

This research aims to normalize quantitative trading in commodities in India, which is currently very limited to a little academic research. As we saw this strategy can have positive returns with a good sharpe ratio and hence can be extended to a corporate setting.

In future research, a variable hedge ratio pairs trading strategy can be used in which the hedge ratio is estimated using a dynamic linear regression model which updates the hedge ratio every day, this can be then incorporated in the spread series and a stochastic model can be applied to it.

\section{References}

\end{document}